\documentclass[3p,times,twocolumn]{elsarticle}

\usepackage{ecrc}

\volume{00}

\firstpage{1}

\journalname{Nuclear Physics B Proceedings Supplement}

\runauth{}

\jid{nuphbp}

\jnltitlelogo{Nuclear Physics B Proceedings Supplement}

\usepackage{amssymb}
\usepackage[figuresright]{rotating}

\newcommand{\MO}{{\tt micrOMEGAs}}
\newcommand{\DS}{{\tt DarkSUSY}}

\def\bpm{\begin{pmatrix}}
\def\epm{\end{pmatrix}}

\def\lr{\left( }
\def\rr{\right) }
\def\le{\left[ }
\def\re{\right] }
\def\lg{\left\{ }
\def\rg{\right\} }

\def\beq{\begin{equation}}
\def\eeq{\end{equation}}
\def\bea{\begin{eqnarray}}
\def\eea{\end{eqnarray}}

\begin{document}

\begin{frontmatter}



\dochead{}

\title{Precision predictions for supersymmetric dark matter}


\author{J. Harz}

\address{Department of Physics and Astronomy, University College London, London WC1E 6BT,
 United Kingdom}

\author{B. Herrmann}

\address{LAPTh, Universit\'e de Savoie, CNRS, 9 Chemin de Bellevue, B.P.\ 110,
 74941 Annecy-le-Vieux, France}

\author{M. Klasen, K. Kovarik, M. Meinecke, P. Steppeler}

\address{Institut f\"ur Theoretische Physik, Westf\"alische
 Wilhelms-Universit\"at M\"unster, Wilhelm-Klemm-Stra\ss{}e 9,
 48149 M\"unster, Germany}

\begin{abstract}
The dark matter relic density has been measured by Planck and its
predecessors with an accuracy of about 2\%. We present theoretical
calculations with the numerical program {\tt DM@NLO} in next-to-leading
order SUSY QCD and
beyond, which allow to reach this precision for gaugino and squark
(co-)annihilations, and use them to scan the phenomenological MSSM
for viable regions, applying also low-energy, electroweak and
hadron collider constraints.
\end{abstract}


\end{frontmatter}

% \linenumbers


\vspace*{-15.5cm}
\noindent LAPTH-Conf-075/14, LCTS/2014-31, MS-TP-14-27
\vspace*{14.1cm}

\section{Introduction}
\label{}

Today, there is ample evidence for the existence of Cold Dark Matter (CDM) in the
Universe from many different scales, ranging from rotational velocity curves
of galaxies and galaxy clusters to graviational lensing, structure formation
and the cosmic microwave background. Through a six-parameter fit to a standard,
spatially flat $\Lambda$CDM cosmological model, the Planck mission has determined
the relic density of the CDM, with an accuracy of about 2\%, to be \cite{Ade:2013zuv}
\beq
 \Omega_{\rm CDM}=0.1199\pm0.0027.
\eeq
Since the Standard Model (SM) of particle physics contains no weakly-interacting,
sufficiently massive particle, this measured value and its uncertainty highly constrain
all SM extensions that could provide a viable dark matter candidate. Among these
extensions, the Minimal Supersymmetric SM (MSSM) has by far received the most attention,
in particular the lightest neutralino ($\tilde\chi_1^0$) stabilised by a discrete ($R$)
symmetry.

Supersymmetry (SUSY) is attractive for many theoretical and phenomenological reasons.
Not only does it relate the two fundamentally different types of particles, fermions
and bosons, through SUSY partners, which otherwise share the same quantum numbers, but
also it represents the maximal possible extension of space-time symmetry (the Poincar\'e
algebra). In addition, the SUSY partners change the $\beta$ functions and thus the
running of couplings and masses, making (in contrast to the SM alone) Grand Unified
Theories (GUTs) phenomenologically viable. Last, but not least, the mass of the lightest,
SM-like Higgs boson $m_{h^0}$ is stabilised, in particular through the contributions of
light top squarks $\tilde t$, which provide a crucial second term in the relation
\bea
 \hspace*{-3mm} m_{h^0}^2\!\!\!&=&\!\!\!m_Z^2\cos^22\beta +{3g^2m_t^4\over8\pi^2 m_W^2}\\
 \hspace*{-3mm} &&\!\!\!\le
 \log{M_{\rm SUSY}^2\over m_t^2}+{X_t^2\over M_{\rm SUSY}^2}\lr
 1-{X_t^2\over12M_{\rm SUSY}^2}\rr\re.\qquad\nonumber
\eea
Here, $X_t=A_t-\mu/\tan\beta$ and $M_{\rm SUSY}=\sqrt{m_{\tilde t_1}m_{\tilde t_2}}$ control
the mixing in the stop sector, which must be nearly maximal ($|X_t|=\sqrt{6}M_{\rm SUSY}$),
with large $A_t$ and a light stop $\tilde t_1$, to explain the relatively high observed
Higgs boson mass of
125.36$\pm0.37({\rm stat.})\pm0.18({\rm syst.})$ GeV \cite{Aad:2014aba} and
125.03$^{+0.26}_{-0.27}({\rm stat.})^{+0.13}_{-0.15}({\rm syst.})$ GeV \cite{CMS:2014ega}, respectively.
Since SUSY spectrum generators such as {\tt SPheno} \cite{Porod:2011nf}
have relatively large theoretical uncertainties, one usually relaxes the Higgs mass
constraint, e.g.\ to 122 GeV $\leq m_{h^0}\leq$ 128 GeV. Other important constraints on
the MSSM model paramters come from loop contributions in rare $B$ meson decays, in
particular $2.77\cdot10^{-4}\leq{\rm BR}(b\to s\gamma)\leq4.33\cdot10^{-4}$ at 3$\sigma$
\cite{HFAG},
and from direct SUSY particle searches at the LHC, with null results in particular for gluinos
and first and second generation squarks with masses below about 1 TeV
\cite{Aad:2014wea,Chatrchyan:2014lfa}.
Currently less
important are constraints from the anomalous magnetic moment of the muon and electroweak
precision measurements at LEP.

All of these constraints are applied in our random scans of a phenomenologically
motivated subset of eight MSSM parameters at the TeV scale (pMSSM-8), e.g.\
\begin{eqnarray}
 500~{\rm GeV} \leq M_{\tilde{q}_{1,2}} &\leq&~ 4000~{\rm GeV}, \nonumber\\
 100~{\rm GeV} \leq M_{\tilde{q}_3}     &\leq&~ 2500~{\rm GeV} , \nonumber\\
 500~{\rm GeV} \leq M_{\tilde{\ell}}   ~&\leq&~ 4000~{\rm GeV}, \nonumber \\
                    |T_t|              ~&\leq&~ 5000~{\rm GeV} ,  \nonumber\\
 200~{\rm GeV} \leq M_1                ~&\leq&~ 1000~{\rm GeV}, \\
 100~{\rm GeV} \leq m_A                ~&\leq&~ 2000~{\rm GeV} ,  \nonumber\\
                            |\mu|      ~&\leq&~ 3000~{\rm GeV} , \nonumber \\
 2             \leq        \tan\beta    &\leq&  50.  \nonumber
\end{eqnarray}
In addition, the GUT relations for $M_{2,3}$ can be relaxed, as can be the parameter $M_{\tilde u_3}$,
leading to eleven free parameters (pMSSM-11).

\section{Neutralino/chargino coannihilation}
\label{}

\begin{figure}
 \centering
 \includegraphics[width=\columnwidth]{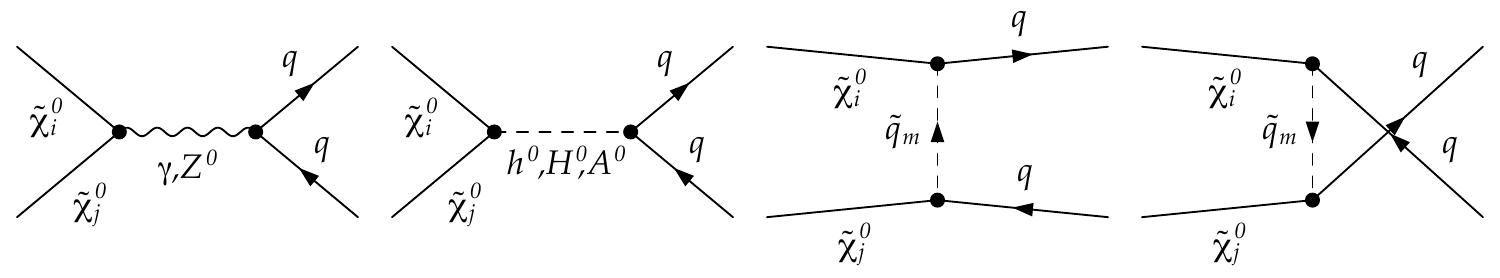}
 \includegraphics[width=\columnwidth]{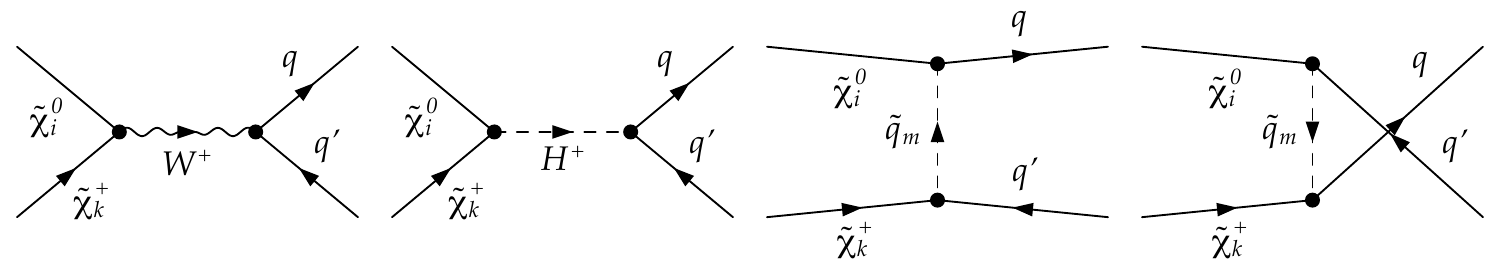}
 \includegraphics[width=\columnwidth]{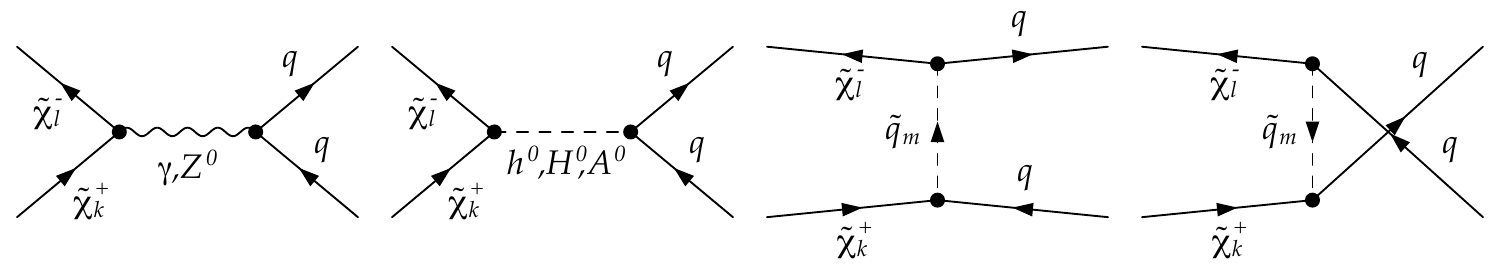}
 \caption{\label{fig:1}Tree level diagrams of the gaugino (co-)annihilation processes
 $\tilde{\chi}_i^0 \tilde{\chi}_j^0 \rightarrow q\bar{q}$ (top), $\tilde{\chi}_i^0
 \tilde{\chi}_k^\pm \rightarrow q\bar{q}'$ (middle), and $\tilde{\chi}_k^\pm \tilde{\chi}_l^\pm
 \rightarrow q\bar{q}$ (bottom).}
\end{figure}

Apart from the annihilation of the CDM particle, the lightest neutralino $\tilde\chi_1^0$,
with itself, many co-annihilation processes can become relevant or even dominant in different
regions of the
pMSSM. These processes may in particular be co-annihilations between different types of
neutral gauginos and higgsinos (neutralinos) and charged gauginos and higgsinos (charginos),
as shown at the tree-level in Fig.\ \ref{fig:1}.
%
%%%%%%%%%%%%%% Begin Figure 1 %%%%%%%%%%%%%%%%%%%%%%%%%%%%%%%%%%%%%%%%%%
%%%%%%%%%%%%%% End of Figure 1 %%%%%%%%%%%%%%%%%%%%%%%%%%%%%%%%%%%%%%%%%
%
%
%%%%%%%%%%%%%% Begin Table 1 %%%%%%%%%%%%%%%%%%%%%%%%%%%%%%%%%%%%%%%%%%%
\begin{table}[h]
\caption{\label{tab:1}Most relevant gaugino (co-)annihilation channels into quarks in three
 typical pMSSM-11 scenarios \cite{Herrmann:2014kma}. Channels which contribute less than 0.1\%
 to the thermally averaged cross section are not shown.}
\begin{center}
\begin{tabular}{|rl|ccc|}
		\hline
		 & & Scen.\ I & Scen.\ II & Scen.\ III\\
		\hline
		$\tilde{\chi}^0_1 \tilde{\chi}^0_1 \to$ & $t\bar{t}$ & 1.4\% & 15.0\% & -- \\
		                                        & $b\bar{b}$ & 9.1\% &  5.9\% & -- \\
		                                        & $c\bar{c}$ & --    &  0.1\% & -- \\
		                                        & $u\bar{u}$ & --    &  0.1\% & -- \\
		\hline
		$\tilde{\chi}^0_1 \tilde{\chi}^0_2 \to$ & $t\bar{t}$ &  2.5\% & 12.0\% & 3.3\% \\
		                                        & $b\bar{b}$ & 23.0\% &  6.9\% & 1.6\% \\
		                                        & $c\bar{c}$ & --     & --     & 1.3\% \\
		                                        & $s\bar{s}$ & --     & --     & 1.7\% \\
		                                        & $u\bar{u}$ & --     & --     & 1.3\% \\
		                                        & $d\bar{d}$ & --     & --     & 1.7\% \\
		\hline
		$\tilde{\chi}^0_1 \tilde{\chi}^0_3 \to$ & $t\bar{t}$ & --  &  9.1\% & -- \\
		                                        & $b\bar{b}$ & --  &  5.3\% & -- \\
		\hline
		$\tilde{\chi}^0_2 \tilde{\chi}^0_2 \to$ & $b\bar{b}$ & 0.2\% & -- & -- \\
		\hline
		$\tilde{\chi}^0_1 \tilde{\chi}^{\pm}_1 \to$ & $t\bar{b}$ & 43.0\% & 40.0\% & 0.8\% \\
													& $c\bar{s}$ & --     & --     & 8.5\% \\
													& $u\bar{d}$ & --     & --     & 8.5\% \\
		\hline
		$\tilde{\chi}^0_2 \tilde{\chi}^{\pm}_1 \to$ & $t\bar{b}$ & 0.4\% & -- & 0.4\% \\
		                                        	& $c\bar{s}$ & 0.9\% & -- & 4.6\% \\
		                                        	& $u\bar{d}$ & 0.9\% & -- & 4.6\% \\
		\hline
		$\tilde{\chi}^{\pm}_1 \tilde{\chi}^{\pm}_1 \to$ & $t\bar{t}$ & 0.2\% & -- & 3.2\% \\
		                                        		& $b\bar{b}$ & 0.6\% & -- & 2.7\% \\
		                                        		& $c\bar{c}$ & 0.2\% & -- & 2.3\% \\
		                                        		& $s\bar{s}$ & 0.2\% & -- & 1.4\% \\
		                                        		& $u\bar{u}$ & 0.2\% & -- & 2.3\% \\
		                                        		& $d\bar{d}$ & 0.2\% & -- & 1.4\% \\
		\hline
		\multicolumn{2}{|c|}{Total} & 83.0\% & 94.4\% & 51.6\% \\
		\hline
\end{tabular}
\end{center}
\end{table}
%%%%%%%%%%%%%% End of Table 1 %%%%%%%%%%%%%%%%%%%%%%%%%%%%%%%%%%%%%%%%%%
%
At this level, all of these processes are routinely included in programs such as
\MO\ \cite{Belanger:2013oya} and \DS\ \cite{Gondolo:2004sc}. They become relevant,
when the ratios of equilibrium densities
\beq
 {n_i^{\rm eq.}\over n_j^{\rm eq.}}\sim \exp\lg-{m_i-m_j\over T}\rg
\eeq
are only weakly Boltzmann suppressed, i.e.\ when the mass differences of the
SUSY particles $i$ and $j$ are small. For three different reference scenarios
\cite{Herrmann:2014kma}, the most relevant gaugino (co-)annihilation contributions
into quarks are shown in Tab.\ \ref{tab:1}.

After generalising these processes to allow for general types of flavour violation
\cite{Bozzi:2007me,Fuks:2008ab,Fuks:2011dg,Herrmann:2011xe}, we have now computed all
their SUSY-QCD corrections and implemented these in a numerical code named
{\tt DM@NLO} ({\tt http://dmnlo.hepforge.org}) \cite{Herrmann:2014kma}.
This includes previously calculated corrections to the annihilation of two lightest
neutralinos into heavy quarks through the Higgs funnel $\tilde\chi_1^0\tilde\chi_1^0\to
A^0\to b\bar{b}$ \cite{Herrmann:2007ku}, other Higgs resonances \cite{Herrmann:2009wk},
or gauge boson and squark exchanges with bottom and top quark final states
\cite{Herrmann:2009mp}.
New in our recent work is not only the generalisation of the initial state,
but also the generalisation of the final state to first and second generation quarks,
which are in general not negligible as one can see from Tab.\ \ref{tab:1}. 

\begin{figure}
 \centering
 \includegraphics[width=\columnwidth]{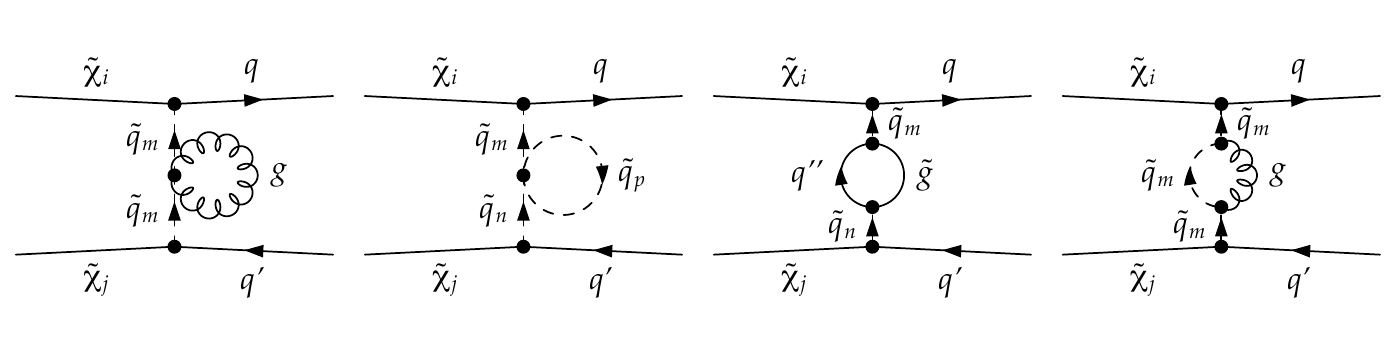}
 \includegraphics[width=\columnwidth]{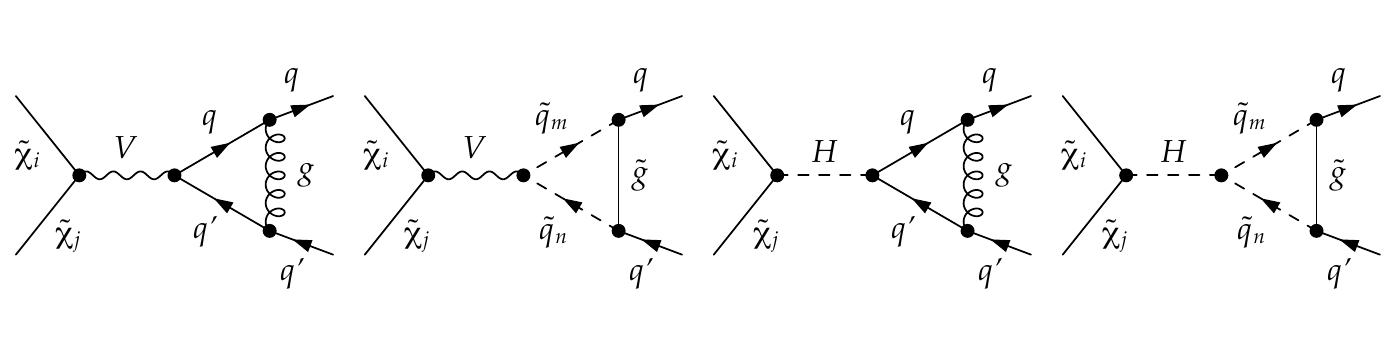}
 \includegraphics[width=\columnwidth]{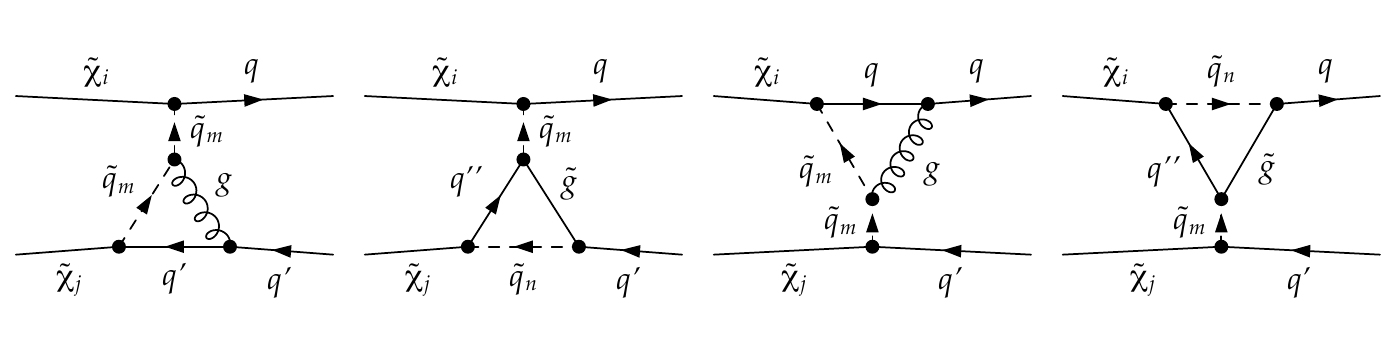}
 \includegraphics[width=\columnwidth]{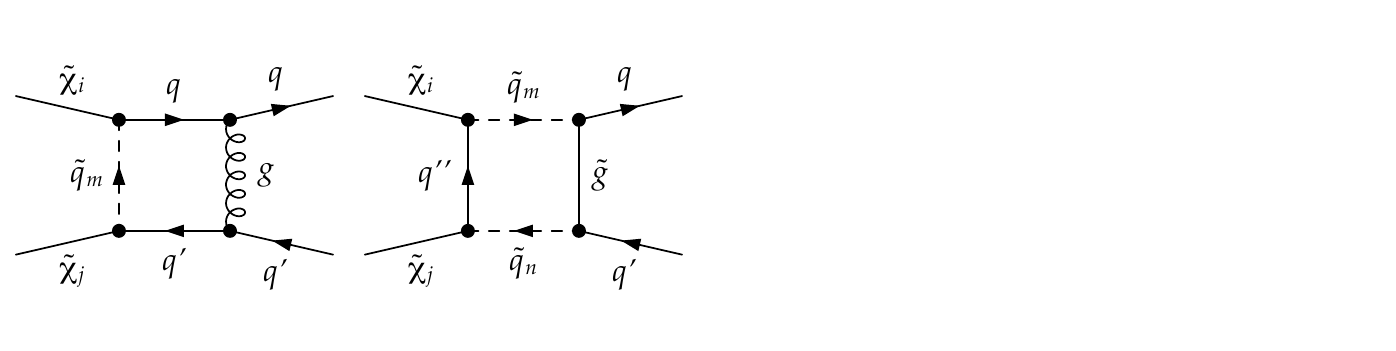}
 \caption{\label{fig:2}Diagrams depicting schematically the one-loop corrections of
 ${\cal O}(\alpha_s)$ to the gaugino (co-)annihilation processes shown in Fig.\ \ref{fig:1}.
 Here, $V = \gamma, Z^0, W^{\pm}$ and $H = h^0, H^0, A^0, H^{\pm}$. The corrections to the
 $u$-channel processes are not explicitly shown, as they can be obtained by crossing from the
 corresponding $t$-channel diagrams.}
\end{figure}

\begin{figure}
 \centering
 \includegraphics[width=\columnwidth]{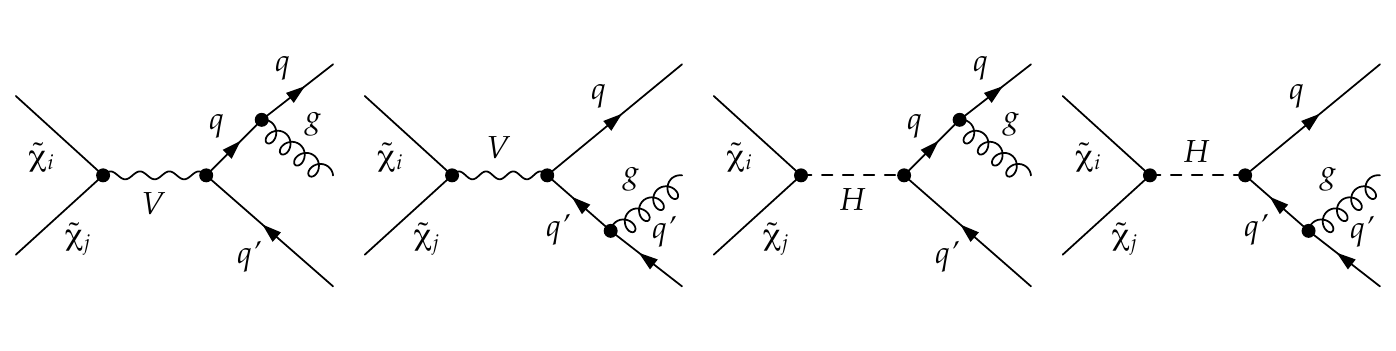}
 \includegraphics[width=\columnwidth]{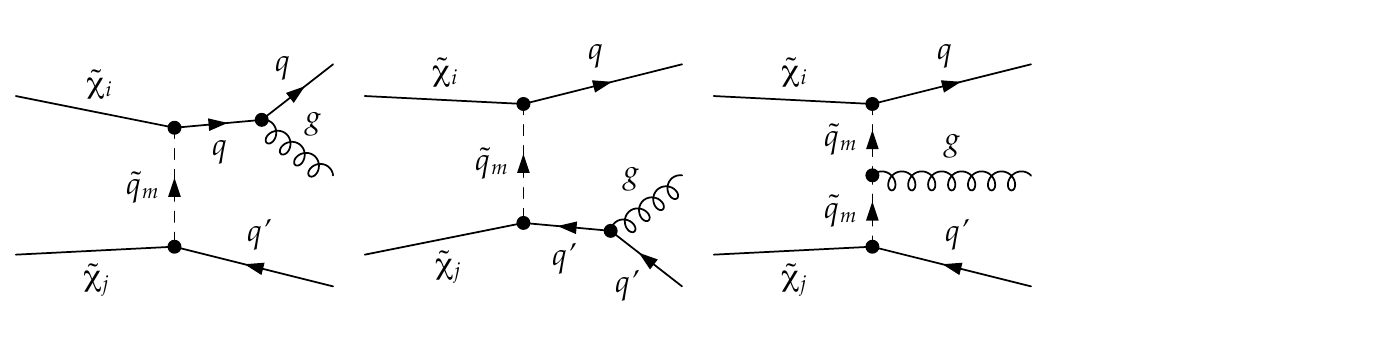}
 \caption{\label{fig:3}Diagrams depicting the real gluon emission corrections of
 ${\cal O}(\alpha_s)$ to the gaugino (co-)annihilation processes shown in Fig.\ \ref{fig:1}.
 As before, $V = \gamma, Z^0, W^{\pm}$ and $H = h^0, H^0, A^0, H^{\pm}$. The corrections to the
 $u$-channel processes are not explicitly shown, as they can be obtained by crossing from the
 corresponding $t$-channel diagrams.}
\end{figure}

Typical one-loop diagrams are depicted in Fig.\ \ref{fig:2},
%
%%%%%%%%%%%%%% Begin Figure 2 %%%%%%%%%%%%%%%%%%%%%%%%%%%%%%%%%%%%%%%%%%
%%%%%%%%%%%%%% End of Figure 2 %%%%%%%%%%%%%%%%%%%%%%%%%%%%%%%%%%%%%%%%%
%
real emission diagrams in Fig.\ \ref{fig:3}.
%
%%%%%%%%%%%%%% Begin Figure 3 %%%%%%%%%%%%%%%%%%%%%%%%%%%%%%%%%%%%%%%%%%
%%%%%%%%%%%%%% End of Figure 3 %%%%%%%%%%%%%%%%%%%%%%%%%%%%%%%%%%%%%%%%%
%
Among them, the infrared poles cancel, while the ultraviolet ones occuring 
in the one-loop diagrams are removed by renormalisation (for details see Ref.\
\cite{Harz:2012fz}). The infrared poles in the real corrections are made
explicit and the finite remainder is treated numerically with the Catani-Seymour
dipole subtraction method \cite{Catani:2002hc},
\begin{equation}
 \hspace*{-5mm}
 \sigma_{\rm NLO} \!= \!\! \int_{3} \biggr[ \mathrm{d}\sigma^{\rm R} \Big|_{\epsilon=0} - 	
 \mathrm{d}\sigma^{\rm A} \Big|_{\epsilon=0} \biggr]\! 
 +\!\! \int_2 \! \left[ \mathrm{d}\sigma^{\rm V} + 
 \!\int_1 \! \mathrm{d}\sigma^{\rm A} \right]_{\epsilon=0}\hspace*{-4mm}.\
\end{equation}

\begin{figure}
 \centering
 \includegraphics[width=\columnwidth]{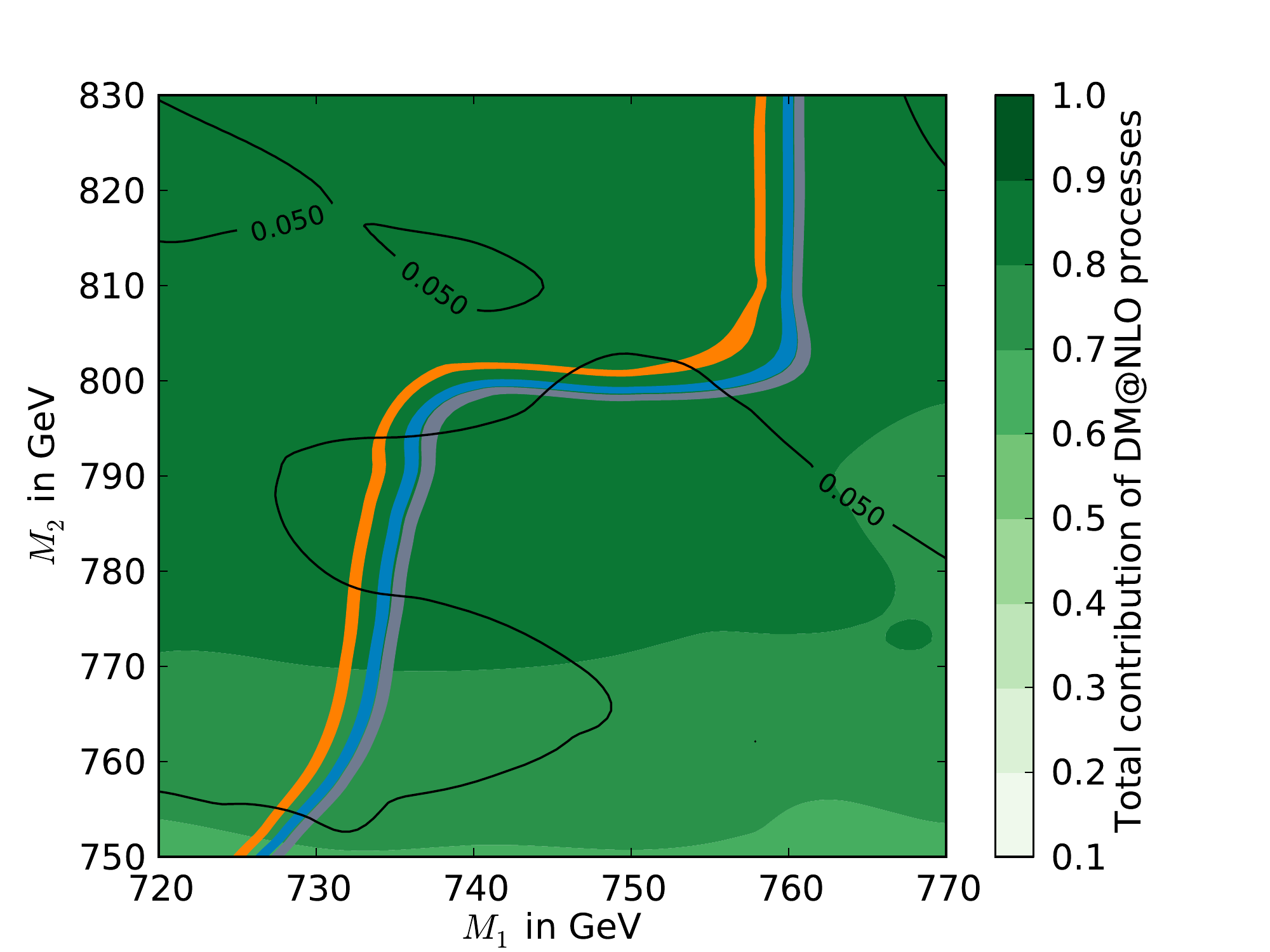}
 \caption{\label{fig:4}Neutralino relic density in the $M_1$--$M_2$-plane surrounding scenario I.
 The three coloured lines represent the part of the parameter space which leads to a neutralino
 relic density compatible with the Planck limits given in Eq.\ (1). For the orange
 line we used the standard \MO\ routine, the grey one corresponds to our tree level calculation,
 and the blue one represents our full one-loop calculation. The black contour lines denote the
 relative shift between the tree level and one-loop relic density, i.e.\
 $\big|1-\Omega_\chi^{\mathrm{NLO}}/\Omega_\chi^{\mathrm{tree}}\big|$.}
\end{figure}

After imposing the relic density constraint, the numerical impact of the corrections
on the viable parameter space becomes visible in Fig.\ \ref{fig:4}.
%
%%%%%%%%%%%%%% Begin Figure 4 %%%%%%%%%%%%%%%%%%%%%%%%%%%%%%%%%%%%%%%%%%
%%%%%%%%%%%%%% End of Figure 4 %%%%%%%%%%%%%%%%%%%%%%%%%%%%%%%%%%%%%%%%%
%
Here we only show the plane of gaugino mass parameters $M_1$ and $M_2$ and fix
all other pMSSM-11 parameters. More than 70 or 80\% of all contributing (co-)annihilation
processes are computed at next-to-leading order (NLO) of SUSY-QCD with {\tt DM@NLO},
leading to corrections to the relic density of 5\% or more and a clearly visible
shift of the Planck contour from tree-level (grey) to NLO (blue). The standard \MO\ 
result, which tries to capture some corrections through effective vertices, agrees
with neither prediction in this particular case.

\section{Electroweak stop coannihilation}
\label{}

\begin{figure}
 \centering
 \includegraphics[width=\columnwidth]{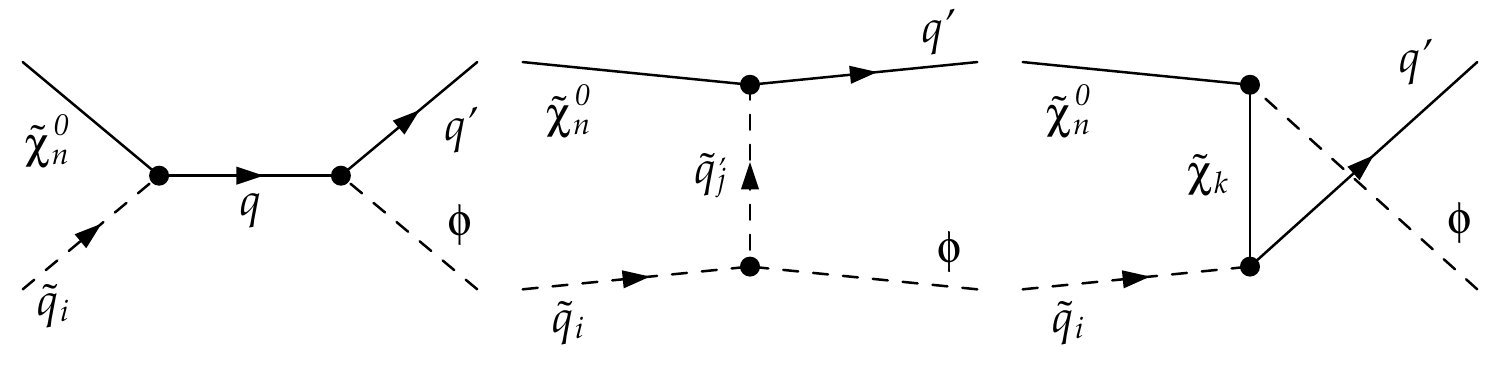}
 \includegraphics[width=\columnwidth]{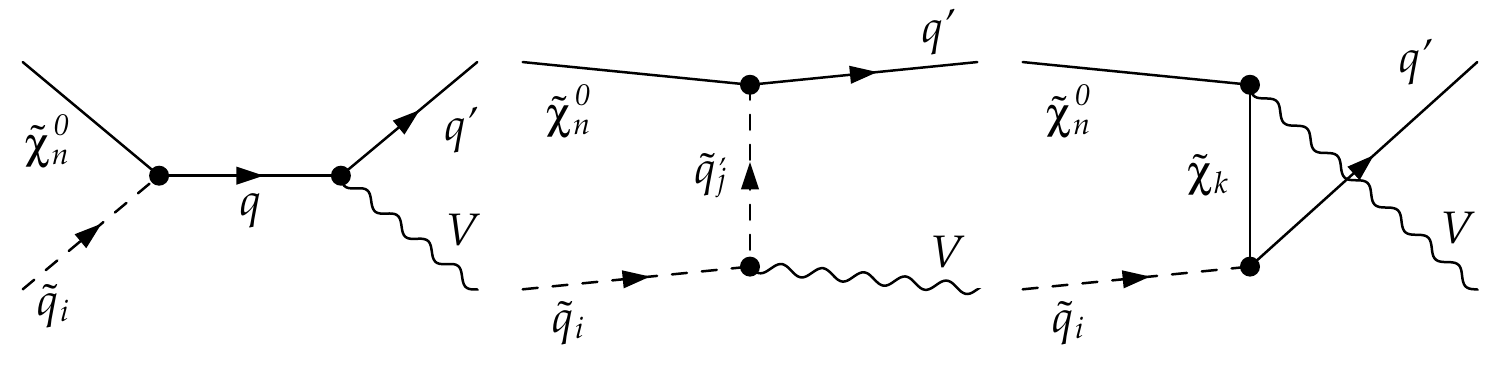}
 \caption{\label{fig:5}Leading-order Feynman diagrams for neutralino-squark co-annihilation
 into a quark and a Higgs boson ($\phi=h^0,H^0,A^0,H^{\pm}$) or an electroweak gauge boson
 ($V=\gamma,Z^0,W^{\pm}$). The $u$-channel is absent for a photon in the final state.}
\end{figure}

\begin{figure}
 \centering
 \includegraphics[width=\columnwidth]{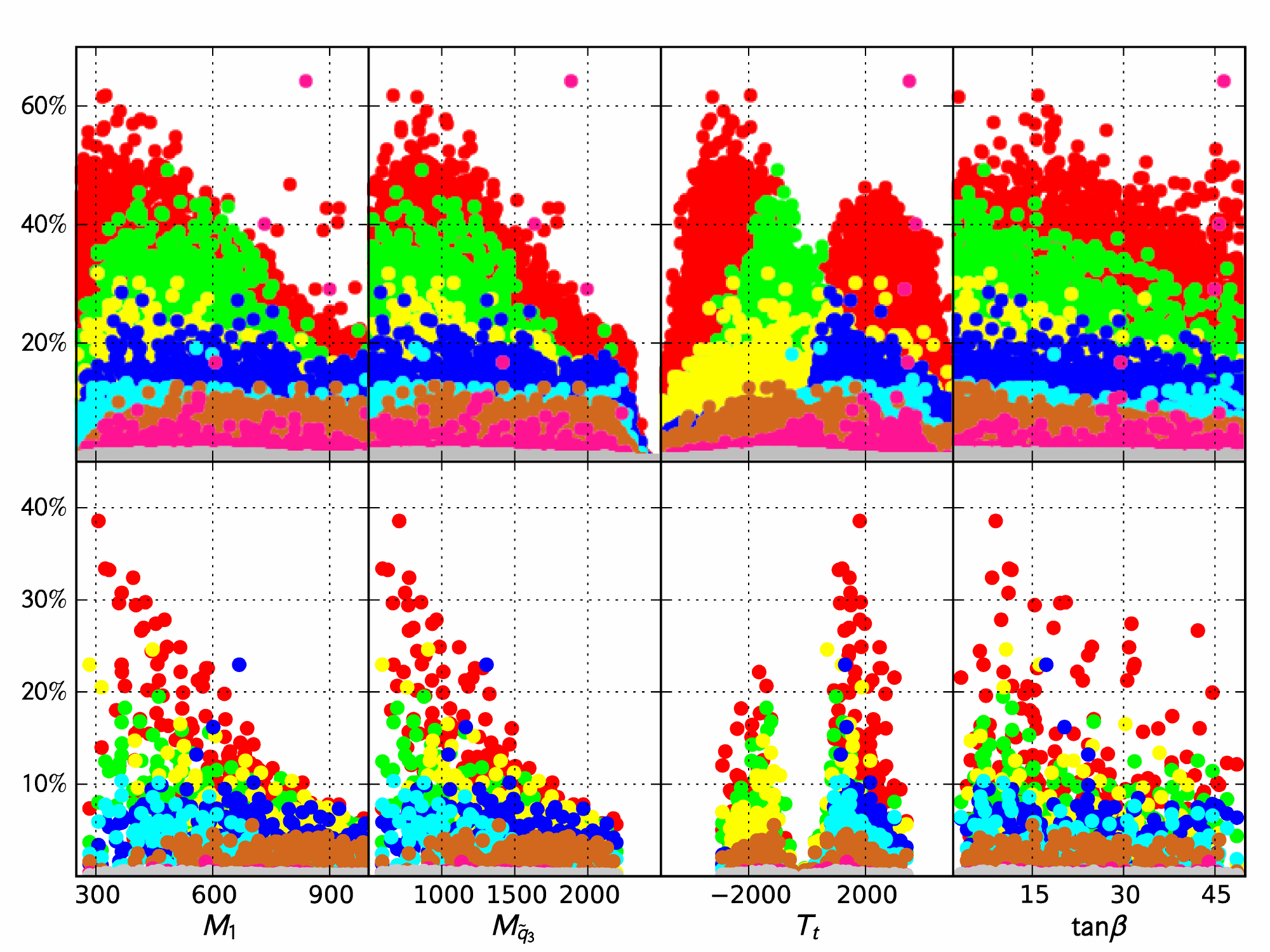}
 \caption{\label{fig:6}Relative contributions of the neutralino-stop co-annihilation
 channels for the generated parameter points as a function of the input
 parameters $M_1$, $M_{\tilde{q}_3}$, $T_t$, and $\tan\beta$ before (top) and
 after (bottom) applying selection cuts.
 Shown are the contributions from $th^0$ (red), $tg$ (green), $tZ^0$ (blue),
 $tH^0$ (yellow), $bW^+$ (cyan), $tA^0$ (chocolate), $bH^+$ (pink), and
 $t\gamma$ (silver) final states. The parameters $M_1$, $M_{\tilde{q}_3}$, and
 $T_t$ are given in GeV.}
\end{figure}

When the top squark mass comes close to the lightest neutralino mass,
as it is natural from the arguments given in the introduction, but also
for cosmological reasons, co-annihilations among these particles may
also occur, as shown at the tree level in Fig.\ \ref{fig:5}.
%
%%%%%%%%%%%%%% Begin Figure 5 %%%%%%%%%%%%%%%%%%%%%%%%%%%%%%%%%%%%%%%%%%
%%%%%%%%%%%%%% End of Figure 5 %%%%%%%%%%%%%%%%%%%%%%%%%%%%%%%%%%%%%%%%%
%
These processes must then also be corrected in a similar way as described
above. This has now been achieved and implemented in {\tt DM@NLO} \cite{Harz:2012fz,Harz:2013aua}.
In a previously performed caluclation \cite{Freitas:2007sa}, only the co-annihilation
of bino-like neutralinos with right-handed stops into top quarks and gluons as well
as bottom-quarks and $W$-bosons was considered. We have now corrected all co-annihilation
processes, irrespective of the gaugino/higgsino or top squark decomposition, and included
all electroweak final states (see Fig.\ \ref{fig:6}).
%
%%%%%%%%%%%%%% Begin Figure 6 %%%%%%%%%%%%%%%%%%%%%%%%%%%%%%%%%%%%%%%%%%
%%%%%%%%%%%%%% End of Figure 6 %%%%%%%%%%%%%%%%%%%%%%%%%%%%%%%%%%%%%%%%%
%
%
%%%%%%%%%%%%%% Begin Figure 7 %%%%%%%%%%%%%%%%%%%%%%%%%%%%%%%%%%%%%%%%%%
\begin{figure}
 \centering
 \includegraphics[width=\columnwidth]{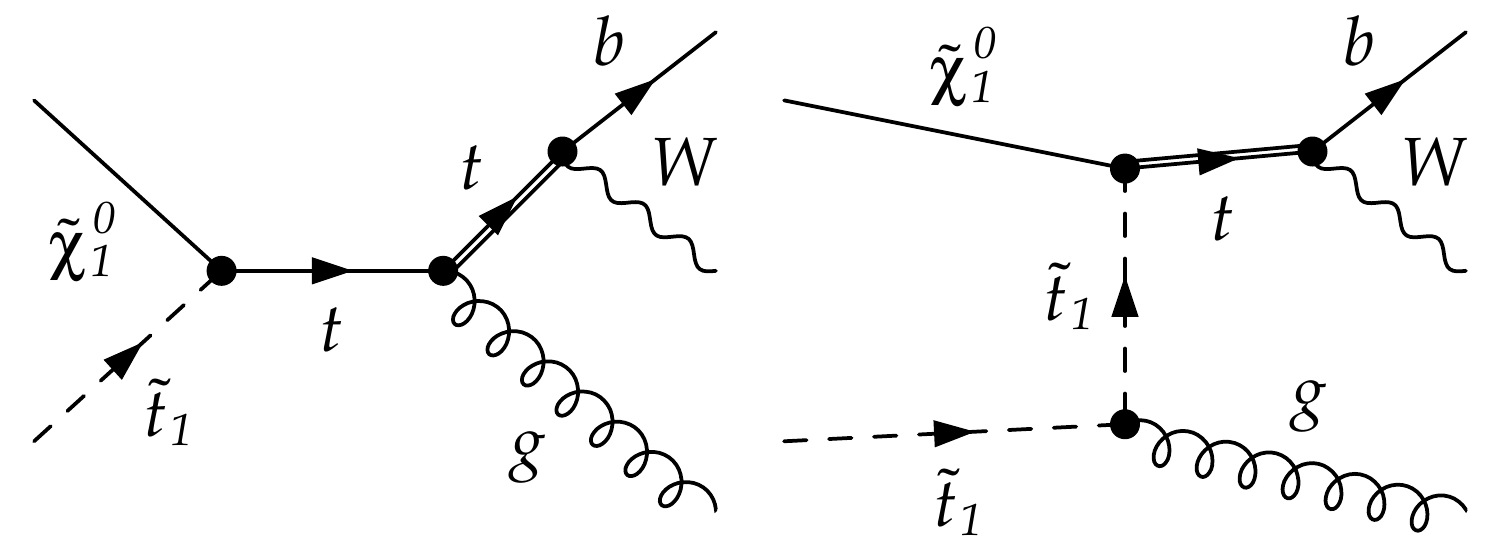}
 \caption{\label{fig:7}Real gluon emission diagrams with a $Wb$ final state where a internal
 top quark can become on-shell, as indicated by a double line.}
\end{figure}
%%%%%%%%%%%%%% End of Figure 7 %%%%%%%%%%%%%%%%%%%%%%%%%%%%%%%%%%%%%%%%%
%
An important subtlety arises from intermediate on-shell top quarks, cf.\ Fig.\ 
\ref{fig:7}. These diagrams also contribute to the electroweak corrections of
top-gluon final states. In order to avoid double counting, we subtract a local
gauge-invariant term defined by the squared resonant amplitude with the top quark
on shell except for the propagator denominator, which is kept as a general
Breit-Wigner function
\begin{equation}
	\left| \mathcal{M}_{2\to 3}^{\rm sub} \right|^2 = 
		\frac{m_t^2 \Gamma_t^2} {(p_t^2-m_t^2)^2 + m_t^2 \Gamma_t^2} 
		\left| \mathcal{M}_{2\to 3}^{\rm res} \right|^2_{p_t^2=m_t^2}
\end{equation}
(cf.\ also Ref.\ \cite{Beenakker:1996ch}).

The numerical impact of our SUSY-QCD corrections on the velocity-weighted
cross section $\sigma v$ 
%
%%%%%%%%%%%%%% Begin Figure 8 %%%%%%%%%%%%%%%%%%%%%%%%%%%%%%%%%%%%%%%%%%
\begin{figure}
 \centering
 \includegraphics[width=\columnwidth]{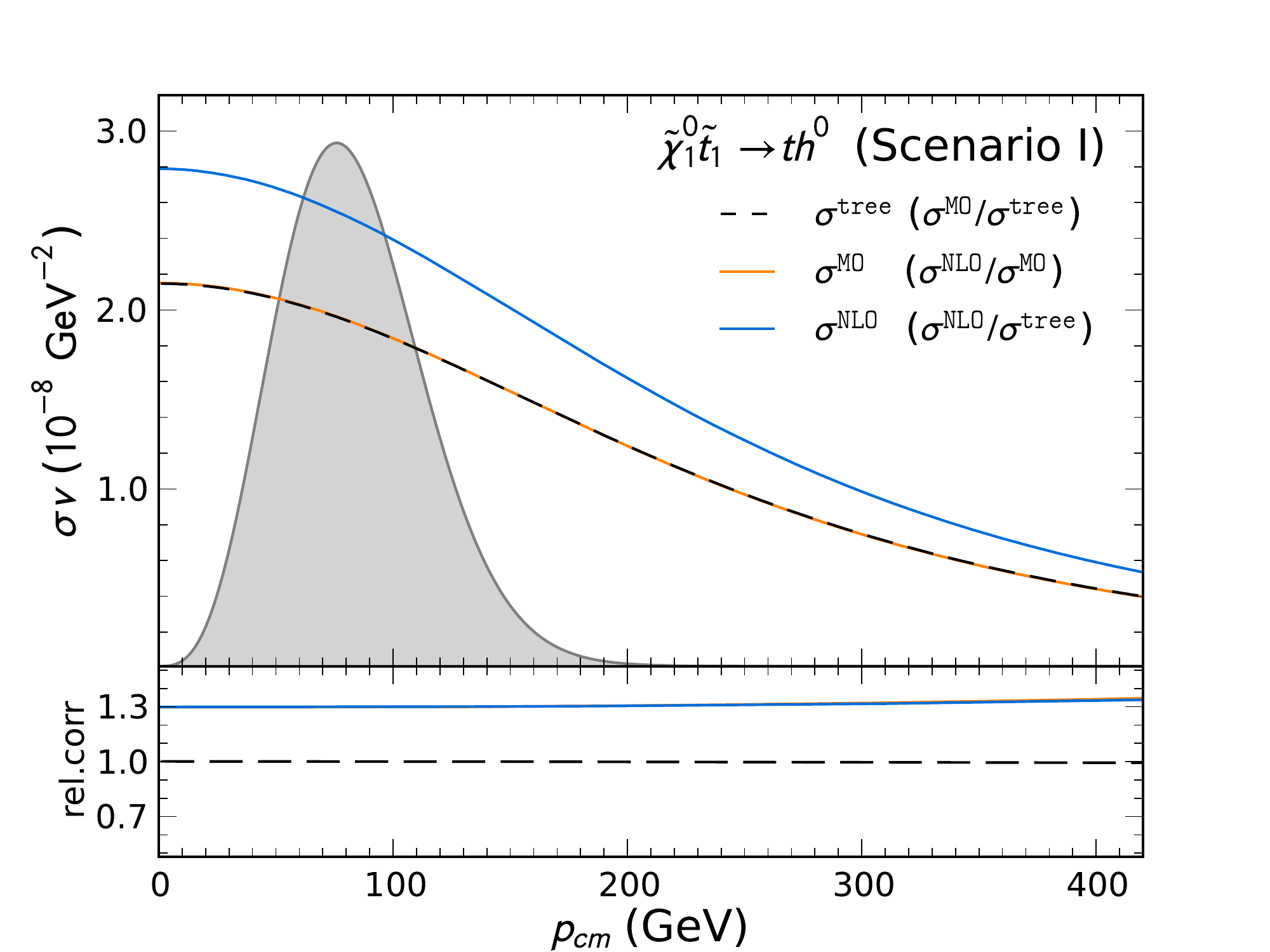}
 \caption{\label{fig:8}Tree-level (black dashed line), full one-loop (blue solid line) and {\MO}
 (orange solid line) cross-section for a top-Higgs final state in a typical reference scenario.
 The upper part of the plot shows the absolute value of $\sigma v$
 together with the thermal distribution (in arbitrary units), whereas the lower part shows the
 corresponding relative shifts (second item in the legend).}
\end{figure}
%%%%%%%%%%%%%% End of Figure 8 %%%%%%%%%%%%%%%%%%%%%%%%%%%%%%%%%%%%%%%%%
%
for a top-Higgs final state in a typical co-annihilation scenario can be seen
in Fig.\ \ref{fig:8}. In this case, the \MO\ prediction (orange) fully agrees
with our tree-level result (grey) when using the same top quark mass, while the
NLO prediction (blue) is about 30\% larger. The relic density is then increased
by about 9\%, which is comparable to the corrections obtained in the gaugino
co-annihilation scenarios.

\section{Conclusion}
\label{}

To summarise, the SUSY-QCD corrections to all gaugino co-annihilation processes
into light and heavy quarks as well as to gaugino-squark co-annihilations into
electroweak final states have now been computed and implemented in {\tt DM@NLO}.
They amount typically to increases in the cross section of 5$-$30\% and shifts
in the Planck relic density band that can exceed its experimental uncertainty.
The effective vertex approach implemented e.g.\ in \MO\ is in general not able
to correctly capture the higher-order corrections, which can induce significant
shifts in the extracted SUSY parameters. Our corrections are implemented using
general couplings, making them generalisable to other, in particular
non-supersymmetric models. Note that loop corrections and co-annihilation processes
can sometimes change the picture even qualitatively, not only quantitatively,
as it has been shown for inert Higgs doublet \cite{Klasen:2013btp} and radiative seesaw
models \cite{Klasen:2013jpa}.

\section*{Acknowledgments}
\label{}

We thank Q.\ Le Boulc'h for his collaboration on the stop co-annihilation channels.
The work of J.H.\ was supported by the London Centre for Terauniverse Studies (LCTS),
using funding from the European Research Council via the Advanced Investigator Grant 267352.
The group in M\"unster is supported by the Helmholtz Alliance for Astroparticle Physics and
the Deutsche Forschungsgemeinschaft under grant KL 1266/5-1.



%% References

%\nocite{*}
%\bibliographystyle{elsarticle-num}
%\bibliography{martin}



\end{document}